\documentclass[12pt,epsf]{article}
\usepackage{graphicx, amsmath}

\begin{document}
\sloppy
\begin{flushright}{SIT-HEP/TM-58}
\end{flushright}
\vskip 1.5 truecm
\centerline{\large{\bf A new perspective on supersymmetric inflation}}
\vskip .75 truecm
\centerline{\bf Tomohiro Matsuda\footnote{matsuda@sit.ac.jp}}
\vskip .4 truecm
\centerline {\it Laboratory of Physics, Saitama Institute of Technology,}
\centerline {\it Fusaiji, Okabe-machi, Saitama 369-0293, 
Japan}
\vskip 1. truecm

\makeatletter
\@addtoreset{equation}{section}
\def\theequation{\thesection.\arabic{equation}}
\makeatother
\vskip 1. truecm

\begin{abstract}
\hspace*{\parindent}
We consider supersymmetric inflation with the hybrid-type potential.
In the absence of the symmetry that forbids Hubble-induced mass terms,
 the inflaton 
 mass will be as large as the Hubble scale during inflation.
We consider gravitational decay of the trigger field as the least decay
mode
 and find that the damping caused by the dissipation can dominate the
 friction of the inflaton when the heavy trigger field is coupled to the
 inflaton. 
The dissipative damping provides a solution to the traditional 
$\eta$ problem without introducing additional symmetry and interactions.
Considering the spatial inhomogeneities of the dissipative coefficient,
we find that modulated inflation (modulation of the inflaton velocity)
can create significant curvature perturbations. 
\end{abstract}
\newpage
\section{Introduction and model}
Warm inflation is known as the scenario of dissipative inflaton
motion\cite{Warm-orig}. 
In this paper we address the generic situation where the inflaton field,
which couples to a massive trigger field, dissipates its energy into
light fields during inflation {\bf without}
 introducing additional interactions.
Namely, a trigger field $\chi$ coupled to the inflaton $\phi$ with the
interaction $\sim \frac{g^2}{2} \phi^2 \chi^2$ obtains large
mass $m_\chi\sim g \phi$ during inflation, and decays (at least
gravitationally) into light matter with a decay rate 
\begin{equation}
\Gamma_{\chi}\simeq \frac{m_\chi^3}{M_p^2}.
\end{equation}
The condition for fast decay ($\Gamma_\chi>H$) is
trivial. 
We define the critical value $\phi_*$ by
\begin{equation}
\phi_* \equiv \frac{\left(H M_p^2\right)^{1/3}}{g}
\end{equation}
and find that the decay into light matter ($\chi\rightarrow 2\psi$)
 is efficient for $\phi>\phi_*$.
For a more concrete argument, we consider the hybrid-type potential
with the superpotential 
\begin{equation}
W=g\Phi(Q_{+}Q_{-}-M^2),
\end{equation}
where $\Phi$ is the inflaton, and $Q_{\pm}$ are a pair of charged
fields (trigger fields) that are responsible for reheating after
inflation.
For the region $\phi\gg M$, we consider the
typical flat potential
\begin{equation}
V(\phi)=V_I+ \frac{c_H}{2} H^2 |\phi|^2 
+\left(\frac{\lambda_n H}{nM_p^{n-3}}\phi^n +h.c.
\right)+|\lambda_n|^2 \frac{|\phi|^{2n-2}}{M_p^{2n-6}},
\end{equation}
where $c_H$ and $\lambda_n$ are dimensionless parameters and
$n$ is an integer greater than 4. 
The first term $V_I\sim g^2M^4$ denotes the inflaton potential caused by
the trigger field, which is displaced from the true minimum to the
origin.
The terms proportional to $\lambda_n$ originate from the typical
supergravity action. 
For simplicity, we consider the trigger field $\chi$, the typical
interaction given by
\begin{equation}
{\cal L}_I = -\frac{1}{2}g^2 \phi^2 \chi^2,
\end{equation}
and the decay rate $\Gamma_{\chi}=\frac{m_\chi^3}{M_p^2}>H$.

Since the decay into light matter ($\chi\rightarrow 2\psi$) is 
effective during $\phi>\phi_*$, the dissipative coefficient of the
inflaton, which is mediated by the heavy trigger field $\chi$, is given
by\cite{BasteroGil:2004tg} 
\begin{equation}
\Gamma \sim 10^{-2}\left(\frac{m_\chi}{M_p}\right)^2 \phi,
\end{equation}
where $g\sim O(1)$ is assumed.\footnote{For the decay rate
$\Gamma_\chi= k m_\chi$, the dissipative coefficient is
given by $\Gamma \simeq 10^{-2}  g^2 k m_\chi$\cite{BasteroGil:2004tg}.}
The strength of the damping is measured by $r$, given by the dissipative
coefficient and the Hubble parameter: 
\begin{equation}
r\equiv \frac{\Gamma}{3H},
\end{equation}
which can be used to rewrite the field equation of the dissipating 
field as 
\begin{equation}
\ddot{\phi}+3H(1+r)\dot{\phi}+V(\phi,T)_{\phi}=0.
\end{equation}
For the dissipation caused by gravitational decay, the ratio
is given by
\begin{equation}
r \sim 10^{-3}\times c_r \frac{\phi^3}{H M_p^2},
\end{equation}
where $c_r$ is a dimensionless constant.\footnote{Here we consider the
most stringent situation with $c_r\sim 1$. Namely, we do not consider
a situation with many light matter fields, which may enhance the 
decay rate by a factor of
$n> 10^{2}$, where $n$ is the number of light fields. The
Planck-scale suppression can be relaxed if the cut-off scale appears
below $M_p$. These effects may lead to a very large $c_r\gg 1$.}  
Note that for the modest values of $\phi<M_p$ and $c_r\sim 1$, strong
 dissipation is not
possible  if the Hubble parameter is large
($H\ge 10^{-3}M_p$) during inflation.
We thus implicitly assume inflation of the scale $H\ll 10^{-3}M_p$.
From the above equations, we find that strong dissipation defined by
$r>1$ is possible due to gravitational decay $\chi\rightarrow2\psi$.
For $c_r\sim 1$, the condition $r>1$ leads to a strong dissipative 
regime given by 
\begin{equation}
\phi> \phi_c\equiv10 \phi_*.
\end{equation}
Thermalization of the decay product, which is very important in warm
inflation, is not always important in solving the $\eta$
problem.\footnote{In fact, massless gauge field might not exist in the
very early Universe if all the gauge symmetries are completely broken by
the large vacuum expectation values of the unstable flat directions.
In this case, the light scalar field that remain after the symmetry
breaking will be the flaton and its supersymmetric
partner(the light flatino).}   
Namely, if the dissipative coefficient is calculated in the
zero-temperature limit, the dissipation may occur without
thermalization, although inflation is not warm in that case. 
If the dissipation is strong for the gravitational decay, we do not need
 to introduce either additional symmetry that forbids a
$O(H)$ mass or an artificial interaction that leads to very efficient
decay of the trigger field.
To examine more explicitly the slow-roll conditions of dissipative
inflaton, we introduce new slow-roll parameters for the motion,
 different from
the conventional slow-roll parameters.
They are given by
\begin{eqnarray}
\epsilon_w &\equiv& \frac{\epsilon}{(1+r)^2},\nonumber\\
\eta_w &\equiv& \frac{\eta}{(1+r)^2},
\end{eqnarray}
where the usual slow-roll parameters ($\epsilon$ and $\eta$) are defined
by 
\begin{eqnarray}
\epsilon&\equiv& \frac{M_p^2}{2}\left(\frac{V_\phi}{V}\right)^2,
\nonumber\\
\eta &\equiv& M_p^2\frac{V_{\phi\phi}}{V}.
\end{eqnarray}
Here the subscript denotes the derivative with respect to the field.
For warm evolution the third slow-roll parameter can be defined as
\begin{equation}
\beta \equiv \frac{r}{(1+r)^3}\frac{V_\phi}{3H^2}
\frac{\Gamma_\phi}{\Gamma},
\end{equation}
which sometimes raises a trivial condition, depending on the model.
Calculating the slow-roll parameter is straightforward.
If the inflaton is far from the origin, terms proportional to 
$\phi^n$ dominate the potential. 
Then, using $V_I\sim H^2 M_p^2$ during inflation, 
the slow-roll parameters are given by
\begin{eqnarray}
\epsilon_w &\simeq& M_p^2 
\left(\frac{\lambda_n H\phi^{n-1}}{H^2 M_p^{n-1}}\right)^2 
\frac{1}{(1+r)^2}
\sim  
\left(\frac{10^3 \times\lambda_n \phi^{n-4}}{c_r M_p^{n-4}} \right)^2 <1
\\
\eta_w &\simeq& \frac{\lambda_n (n-1)H \phi^{n-2}}{H^2 M_p^{n-3}}
\frac{1}{(1+r)^2}
\sim \frac{10^{6}\times\lambda_n (n-1) H\phi^{n-8}}{c_r^2 M_p^{n-7}}
<1.
\end{eqnarray}
The condition $\epsilon_w<1$ leads to 
\begin{eqnarray}
n=4 & \rightarrow & \lambda_4 <10^{-3} \times c_r\nonumber\\
n>4 &\rightarrow& \phi< \left(10^3 \times \frac{\lambda_n}{c_r} 
\right)^{-1/(n-4)} M_p,
\end{eqnarray}
 and the second condition $\eta_w<1$ leads to 
\begin{eqnarray}
n<8 & \rightarrow & \phi > \phi_\eta^{n<8}
 \equiv \left(10^6 \times \frac{\lambda_n(n-1)
H M_p^{7-n}}{c_r^2}\right)^{1/(8-n)},\\
n=8 &\rightarrow& \lambda_n < 10^{-7} \times \frac{c_r^2 M_p}{H},\\
n>8 &\rightarrow& \phi< \phi_\eta^{n>8} \equiv 
\left(10^{-6}\times 
\frac{c_r^2M_p}{\lambda_n (n-1)H }\right)^{1/(n-8)} M_p.
\end{eqnarray}
The simplest example is $n=4$ and $\lambda_4< 10^{-3}$,
which leads to dissipative slow-roll during $\phi>\phi_\eta$.
We show a schematic picture in Figure 1. 
Introducing $\phi_0$ defined by
\begin{equation}
|\phi_0| \sim \left(\frac{c_H H M_p^{n-3}}{\lambda_n}\right)^{1/(n-2)},
\end{equation}
we find that the above conditions are not applicable for $\phi< \phi_0$,
 since in this region the quadratic term is more important than the
 higher dimensional terms. 
For $\phi<\phi_0$, the slow-roll conditions are given by
\begin{eqnarray}
\epsilon_w &\simeq& M_p^2 
\left(\frac{c_H\phi}{M_p^2}\right)^2 
\frac{1}{(1+r)^2}
\sim  
\left(\frac{10^3 \times c_H H M_p }{c_r \phi^2}\right)^2 <1,
\\
\eta_w &\simeq& \frac{c_H}{(1+r)^2}
\sim \frac{10^{6}\times c_H H^2 M_p^4}{c_r^2\phi^6}<1,
\end{eqnarray}
which lead to
\begin{eqnarray}
\epsilon_w <1 &\rightarrow& \phi> \sqrt{10^3\times 
\frac{c_H H M_p}{c_r}}\nonumber\\
\eta_w<1  &\rightarrow& \phi> 10 \times
\left(\frac{c_H H^2 M_p^4}{c_r^2}\right)^{1/6}.
\end{eqnarray}
We thus find that strong dissipation is indeed possible for 
$\phi< \phi_0$.
However, in contrast to standard hybrid inflation, 
the slow-roll does not end with the waterfall at
$\phi=\phi_e$.

We thus find that slow-roll inflation is possible for gravitational
dissipation,  but the
parameter space for the slow-rolling is not trivial, depending on the
model. 
We show a schematic picture for a successful scenario in Figure 1.
The distinguishing property of this scenario is the appearance of
$\phi_1$ that defines the end of dissipative slow-roll inflation.
Note that in the usual hybrid-type inflation model, the end of inflation
is defined by $\phi_e\equiv M^2/g$, where $\chi=0$ becomes unstable.
In the present model, in contrast to non-dissipative inflation, 
the dissipative slow-roll ends at $\phi=\phi_1$, where the dissipative
slow-roll parameter reaches unity.  
The inflaton motion is fast during $\phi_e<\phi<\phi_1$, which
occurs just before reheating at $\phi=\phi_e$.

\begin{figure}[h]
 \begin{center}
\begin{picture}(200,210)(110,330)
\resizebox{15cm}{!}{\includegraphics{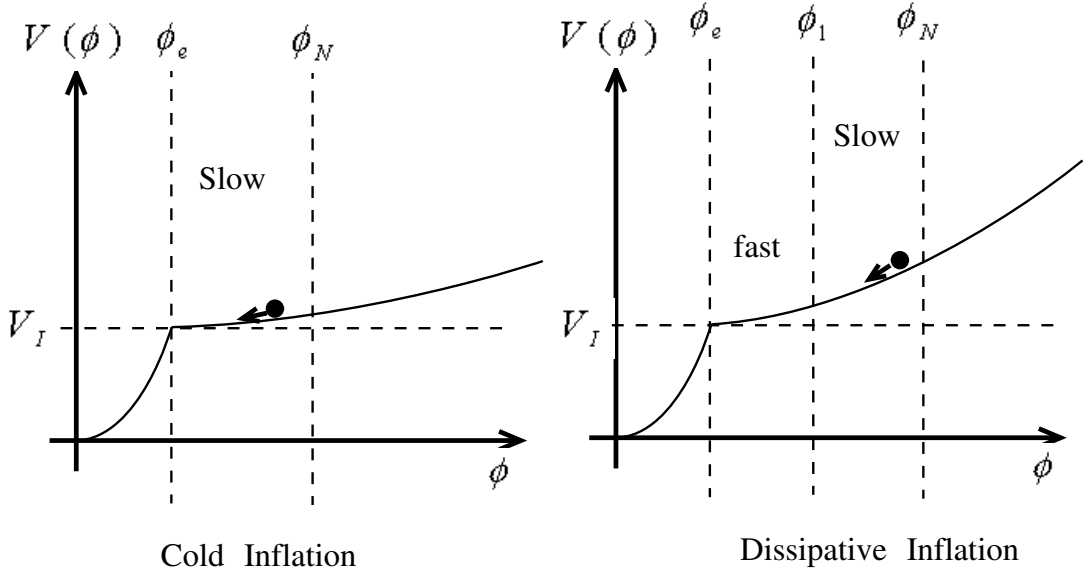}} 
\end{picture}
\label{fig:perspective}
 \caption{$V(\phi)$ represents the typical inflaton potential.
The vacuum energy $V_I$ decays at $\phi=\phi_e$, where the trigger field
starts to roll toward the true minimum.
 $\phi_N$ is defined at the horizon exit for the number of
  e-foldings $N_e$. $\phi_1$ is derived from the dissipative slow-roll
  conditions.} 
 \end{center}
\end{figure}

\section{Curvature perturbations}
In the above argument, we considered the $\eta$-problem in
supersymmetric inflation with the typical hybrid-type potential,
and showed that dissipative slow-roll is possible.
This solves the traditional $\eta$ problem, though 
 it is still not certain if the usual curvature perturbations
created at the horizon exit meet the requirements of the
primordial cosmological perturbations.
In fact, the COBE normalization leads to\cite{EU-books}
\begin{equation}
5.3\times 10^{-4}=\frac{r n V_0^{3/2}M_p^{n-6}}{\lambda_n H \phi_N^{n-1}}
\simeq \frac{10^{-3} \times c_r n H M_p^{n-5}}{\lambda_n \phi_N^{n-4}}
\times P_{bg}(r,r_T),
\end{equation}
where $P_{bg}(r,r_T)$ denotes the possible correction from the background
radiation. 
Of course, $P_{bg}=1$ if there is no significant radiation background
coupled to the inflaton.
Here we have introduced a new parameter defined by $r_T\equiv
T/H>1$.
The source of the radiation is not specified here.
From the Langevin equation, the root-mean square fluctuation
amplitude of the inflaton field $\delta \phi$ after the freeze out 
is obtained to be
\begin{eqnarray}
\delta \phi \equiv \frac{P_{bg}(r,r_T) H}{2\pi} 
&\sim& 
\begin{array}{cccc}
(\Gamma H)^{1/4}T^{1/2}
&\sim& r^{1/4}r_T^{1/2}H 
& (r>1, r_T>1).
\end{array}
\end{eqnarray}
In the above argument, we assumed $\phi_N>\phi_0$.
For $\phi_N<\phi_0$, if the slow-roll conditions are satisfied,
the COBE normalization leads to
\begin{equation}
5.3\times 10^{-4}=\frac{r V_0^{3/2}}{M_p^3 H^2 \phi_N}
\simeq 10^{-3} \times \frac{P_{bg}(r,r_T)\phi_N^{2}}{M_p^2}.
\end{equation}
These results depend crucially on the model.
Notably, we cannot specify the temperature of the background radiation
for the supersymmetric inflation model,
since there are many directions in supersymmetric theory,
which may dissipate and source the background\cite{warm-flat},
even though they are not the inflaton.

The above curvature perturbations, which are calculated using
the standard perturbations created at the horizon crossing, may
 not meet the cosmological requirements.
However, the mismatch does not mean that there is a crucial
difficulty in the scenario.
Namely, alternative scenarios for the curvature perturbations may be 
important for the model, as there are typically many flat directions
in the supersymmetric inflation model.
We thus consider in the next subsection an alternative for the
generation of the curvature perturbations, which works specifically 
in the dissipative scenario.
Other possibilities, such as inhomogeneous scenarios of
phase transition\cite{IH-pt}, (p)reheating\cite{IH-pr,
inhomogeneous-trap} or curvatons\cite{curvatons-original,
curvatons-matsuda} are not discussed in this paper, because they are not 
characteristic of dissipative inflation. 
See Ref.\cite{warm-flat} for more applications of the warm-flat
directions. 

\section{Modulated inflation with inhomogeneous dissipation}
The inflaton potential and the interactions during inflation may be
inhomogeneous due to 
entropy perturbations related to light fields such as moduli or flat
directions in the supersymmetric model\cite{Modulated-kofman}.
The inhomogeneities in the potential or in the interactions 
may lead to inhomogeneous couplings
or masses that may cause fluctuations of the inflaton velocity.
This is the basic idea of modulated inflation\cite{Modulated-matsuda}. 
A similar situation may lead to the inhomogeneous end of hybrid
inflation, as has been discussed in Ref.\cite{End-Modulated}.
However, in the present scenario the inflaton velocity at the end of
inflation is rapid (i.e., $\epsilon_w\sim 1$ or $\eta_w \sim 1$ 
at the end of inflation)
and the perturbations caused by the end-boundary are less significant
than the typical cosmological perturbations.
On the other hand and contrary to the conventional (cold) hybrid model,
the perturbations caused by the modulated inflaton velocity 
($\delta \dot{\phi}$) can be significant in the dissipative inflation
model. 
In this section, we examine the scenario of modulated inflation in
the dissipative inflation model.

For a typical example, we consider the $S$-dependence of the
dissipative coefficient $\Gamma \propto [g(S)]^3$ and the corrections
suppressed by the cut-off scale:
\begin{equation}
g(S) \equiv g_0\left(1+ a_1 \frac{S}{M_*}+...\right),
\end{equation}
where $S$ is a light field that may be warm during
inflation\cite{warm-flat}.
It is possible to assume a hierarchy between the cut-off
scale $M_*$ and the Planck scale ($M_* \ll M_p$).
Note that in the usual cold model, in which the dissipation is
negligible and $\chi=0$ during 
inflation, $\delta g$ cannot cause fluctuations of the inflaton
velocity.
In contrast to the cold model, in the dissipative inflation model
$\delta r/r \ne 0$ caused by  $\delta g$ can lead to
significant inhomogeneities in the inflaton velocity.
Inhomogeneous dissipation ($\delta r/r \ne 0$) creates curvature
perturbations given by\cite{Modulated-matsuda}
\begin{equation}
\delta N \sim \int \frac{\delta(\dot{\phi}^2)}{\dot{\phi}^2}dt
\sim2 N_e \frac{\delta r}{r} 
\sim 6 a_1 N_e \frac{\delta S}{M_*}.
\end{equation}
Here the flat direction $S$ may obtain $O(H)$ mass from the supergravity,
but it still can  move slowly if the dissipation is
significant for the flat direction\cite{warm-flat}. 
Also, for the warm-flat direction $S$, the amplitude of the perturbation
is given by
\begin{eqnarray}
\label{pert-phi}
\delta S &\sim& 
(\Gamma_S H)^{1/4}T^{1/2}
\sim r_S^{1/4}r_T^{1/2}H,
\end{eqnarray}
which shows that the amplitude is characterized by its interaction. 
Here we considered strong dissipation $r_S \equiv \Gamma_S/3H >1$
for the light field $S$.
The amplitude of the warm direction is always larger than the amplitude
of the cold-flat direction. 
The source of the background radiation is not specified, since many
directions in the supersymmetric theory may dissipate during inflation
and contribute to the background radiation.
The spectral index of the curvature perturbation is usually determined
by the potential and the interaction of the light field $S$, but for
dissipative inflation it may also depend on the background radiation
which may be sourced from the other flat directions.

\section{Conclusions and discussion}
In this paper we considered a dissipative inflation model for conventional
supergravity using the typical hybrid-type potential.
If there is no symmetry that forbids Hubble-induced mass, the inflaton
mass will be as large as the Hubble scale during inflation. 
This is the famous $\eta$ problem of the supergravity inflation model.
The purpose of this paper is to find a natural solution to the problem
without introducing symmetries or artificial interactions that may ruin
the simplicity of the original hybrid inflation.
In order to solve the problem within the naive setup, we considered
only the gravitational decay mode (the gravitational decay of the
trigger field into light matter) and find that the dissipation of the
inflaton may be large and can dominate the friction of the inflaton
motion. 
Then we showed that dissipative damping can provide a solution
to the traditional 
$\eta$ problem.
The dissipative inflation scenario is of course not the same as the cold
scenario.
A significant discrepancy that characterizes the dissipative scenario
is the way the inflation ends.
In the usual cold model, inflation ends with the so-called ``waterfall''
of the trigger field at $\phi=\phi_e$, while in the dissipative model
slow-roll ends before the waterfall.
The differences in the dynamics of the inflaton motion lead to 
differences in the inflationary scenario, especially for the generation
of curvature perturbations.
Considering the perturbations of the dissipative coefficient,
we found that modulated inflation can create significant curvature
perturbations.

\section{Acknowledgments}
We wish to thank K.Shima for encouragement, and our colleagues at
Tokyo University for their kind hospitality.

\end{document}